\documentclass[aps,prl,twocolumn,showpacs,preprintnumbers,superscriptaddress]{revtex4}
\usepackage{amssymb}
\usepackage{graphicx}
\usepackage{dcolumn}
\usepackage{bm}
\newcommand{\nc}{\newcommand}
\nc{\be}{\begin{equation}}
\nc{\ee}{\end{equation}}
\nc{\bea}{\begin{eqnarray}}
\nc{\eea}{\end{eqnarray}}
\nc{\bean}{\begin{eqnarray*}}
\nc{\eean}{\end{eqnarray*}}
\nc{\mb}{\mbox}
\nc{\rnc}{\renewcommand}
\nc{\vk}{\mb{\bf k}}
\nc{\vp}{\mb{\bf p}}
\nc{\vn}{\mb{\bf n}}
\nc{\vq}{\mb{\bf q}}
\nc{\rr}{\mb{\bf r}}
\nc{\vz}{\hat {\mb{\bf z}}}
\nc{\vj}{\mb{\boldmath$j$}}
\nc{\vg}{\mb{\boldmath$g$}}
\nc{\x}{\mb{\boldmath$x$}}
\nc{\A}{\mb{\boldmath$A$}}
\nc{\va}{\mb{\boldmath$a$}}
\nc{\vs}{\mb{\boldmath$\sigma$}}
\nc{\vpi}{\mb{\boldmath$\pi$}}
\nc{\nab}{\nabla}
\nc{\X}{\sf x}

\begin{document}

\title{Intra-Landau-Level Cyclotron Resonance in Bilayer Graphene}

\author{Yafis Barlas}
\affiliation{Department of Physics, The University of Texas at Austin, Austin Texas 78712}
\author{R. C\^{o}t\'{e}}
\affiliation{Department of Physics, University of Sherbrooke, Sherbrooke, Quebec, Canada, J1K 2R1}
\author{K. Nomura}
\affiliation{Department of Physics, Tohoku University, Sendai, 980-8578, Japan}
\author{A.H. MacDonald}
\affiliation{Department of Physics, The University of Texas at Austin, Austin Texas 78712}

\begin{abstract}
Interaction driven integer quantum Hall effects are 
anticipated in graphene bilayers because of the near-degeneracy of
the eight Landau levels which appear near the neutral system Fermi level.
We predict that an intra-Landau-level cyclotron resonance signal will appear 
at some odd-integer filling factors, accompanied by collective modes
which are nearly gapless and have approximate $k^{3/2}$ dispersion. 
We speculate on the possibility of unusual localization physics
associated with these modes.
\end{abstract}
\pacs{73.43.-f,76.40.+b,75.30.Ds}

\maketitle


\noindent
{\em Introduction}--- Because the Zeeman spin-splitting in most two-dimensional 
electron systems (2DES's) is much smaller than the Landau level separation, 
the magnetic band spectrum usually consists of narrowly-spaced doublets.
When one of these doublets is 
half-filled and disorder is weak, Coulomb interaction physics leads to ferromagnetism
{\em i.e.} to spontaneous spin polarization in the absence of a Zeeman field \cite{kallinhalperin1984,spinqhf,moonprb}. 
In some circumstances \cite{jungwirth} other approximate Landau level degeneracies occur,
often associated with layer degrees of freedom.  These can also lead to broken symmetries 
which induce quasiparticle gaps and hence interaction driven integer quantum Hall effects. 
The case of bilayer 2DES's is particularly interesting because the {\em which layer} degree of 
freedom doubles Landau level degeneracies and leads to 
exciton condensation \cite{fertigahm,ahmphysicacondferencepaper} at 
odd filling factors and to canted anti-ferromagnetic states \cite{dassarmarajaramanahm} at even
filling factors.  In this Letter, we address the still richer case of graphene bilayer 2DES's 
in which chiral bands lead to an additional degeneracy doubling \cite{graphenebilayertheory}
at the Fermi energy of a neutral system.  Bilayer graphene's Landau level octet is already apparent in present 
experiments \cite{graphenebilayerexpt} through the $8\times(e^2/h)$ Hall conductivity jump between 
well formed plateaus at Landau level filling factors $\nu=-4$ and $\nu=+4$.  We anticipate 
that when external magnetic fields are strong enough or disorder is weak 
enough \cite{macdonaldnomuracriterion}, interactions will drive quantum Hall effects 
at the octet's seven intermediate integer filling factors.  We predict 
that these quantum Hall ferromagnets (QHFs) will exhibit unusual intra-Landau-level cyclotron modes at odd 
filling factors, and that the collective mode excitations at these filling factors are nearly 
gapless even when there is no continuous symmetry breaking.  Because the conductivity has 
Drude weight centered near zero-energy, we speculate that localization physics and quantum-Hall related 
transport phenomena will also be anomalous.  

\noindent
{\em{Graphene Bilayer Landau Levels}}---When trigonal warping \cite{trigonalfootnote} and Zeeman coupling are neglected, 
the low energy properties of Bernal stacked unbalanced bilayer graphene are determined by 
electron-electron interactions and by a band Hamiltonian \cite{graphenebilayertheory}
$\mathcal{H} = \mathcal{H}_{0} + \mathcal{H}_{ext}$ where 
\begin{equation}
\mathcal{H}_{0} =  \frac{1}{2m} \left( \begin{array}{cc} 
0& \pi^{\dagger\,2} \\
\pi^{2} & 0 
\end{array} \right), 
\label{balanced} 
\end{equation}
and the influence of an external potential difference $\Delta_V$ between the layers is captured by  
\begin{equation}
\mathcal{H}_{ext} = \xi \Delta_{V} \left[ \frac{1}{2} \left( \begin{array}{cc} 
1 & 0 \\
0 & -1 
\end{array} \right) - 
\frac{v^{2}}{\gamma_{1}^{2}} \left( \begin{array}{cc} 
\pi^{\dagger}\pi & 0 \\
0 & - \pi \pi^{\dagger} 
\end{array} \right) \right].
\label{unbalanced} 
\end{equation}
In Eqs. (\ref{balanced})-(\ref{unbalanced}), $\vec{\pi} = \vec{p} +(e/c) \vec{A}$ is the 2D kinetic 
momentum, $\pi =\pi_x+i\pi_y$, the $2\times 2$ matrices act on the pseudospin degree of freedom
associated with the two low energy sites \cite{graphenebilayertheory} (the top and bottom layer
sites without a near-neighbor in the opposite layer), $v$ is the single-layer Dirac velocity, 
$\gamma_1 \sim 0.4$eV is the inter-layer hopping amplitude,
and the effective mass $m = \gamma_1/2v^2 \approx 0.054m_{e}$.
$\mathcal{H}$ describes both K ($\xi=1$) and K' ($\xi=-1$) valleys provided that we choose the 
pseudospin representation $(A,\tilde{B})$ for K and $(\tilde{B},A)$ for K'.

Defining the usual raising and lowering Landau level ladder operators $a^{\dagger},a$ with $a^{\dagger} =
(l_{B}/\sqrt{2} \hbar) \pi$,  where $l_{B} = (\hbar c/e B)^{1/2} = 25.6/{\sqrt(B[{\rm Tesla}])}\rm nm$ is the 
magnetic length,
zero-energy eigenstates of $\mathcal{H}_{0}$ can be identified using the property that $a^{2}
\phi_{n} = 0$ for 2D orbitals with Landau level index $n =0,1$.
{\em In bilayer graphene the $n=0$ and $n=1$ orbital Landau levels are members of the same octet.}
This peculiarity is behind most of the physics explored in this paper.
Neutral bilayer graphene's Landau-level octet is the 
direct product of three $S=1/2$ doublets: real spin and which-layer \cite{valleylayercaveat} pseudospins
(as in a normal bilayer), and the Landau-level pseudospin $n=0,1$ degree of freedom which is responsible
for new physics.  Zeeman coupling produces real spin-splitting $\Delta_{Z}$ while   
$\Delta_{V}$ gives rise to layer-splitting as in normal bilayers, but also to a small splitting of the  
Landau-level pseudospin which plays a central role in the physics: 
$ \Delta_{LL} =  \Delta_{V} \hbar \omega / \gamma_{1} \equiv \hbar \omega_{LL}$
where $ \hbar \omega ={ 2\hbar^{2}v^{2}/l_{B}^{2}\gamma_{1}} =  2.14 \, B[\rm Tesla]$ meV.  

%
\noindent
{\em{Octet Hunds Rules}}--- The octet HF Hamiltonian \cite{AllanHF} contains
single-particle pseudospin splitting fields and direct and exchange 
interaction contributions: 
\begin{widetext} 
\begin{equation} 
\label{hfham}
\langle n \tau \alpha | \mathcal{H}_{HF}|n' \sigma  \beta \rangle = E_{H} 
(\rho_{\tau} - \rho_{\beta}) -\sum_{n_{1}n_{2}} 
 \big( X^{+}_{n_{2}n'nn_{1}} + \xi_{\tau} \xi_{\sigma} 
X^{-}_{n_{2}n'nn_{1}} \big) \rho^{n_{1}n_{2}}_{\tau \sigma \alpha \beta} +
(\xi_{\tau} \Delta_{LL} \delta_{n,1} \delta_{n',1}
- \frac{\Delta_{Z}}{2} \xi_{\alpha} \delta_{n n'} -
\frac{\Delta_{V}}{2} \xi_{\tau} \delta_{n n'} ) \delta_{\alpha \beta} \delta_{\tau \sigma},
\end{equation}
\end{widetext}
where $n=0,1$ are LL indices, $\tau, \sigma = t(b) $ are valley indices, $\alpha, \beta = \uparrow(\downarrow) $ are spin-indices,
and $\xi_{\tau(\alpha)} = 1(-1)$ for t(b) layer and $ \uparrow(\downarrow) $ spins respectively. 
In Eq.~(\ref{hfham}) $\rho_{\tau} = \sum_{n\alpha} 
\rho^{nn}_{\tau \tau \alpha \alpha}$ is the total electron density in layer $ \tau$.
The density-matrix $\rho^{n_{1}n_{2}}_{\tau \sigma \alpha \beta} = 
\langle c^{\dagger}_{n_{2} \sigma \beta} c_{n_{1} \tau \alpha} \rangle $ must be determined self consistently 
by occupying the lowest energy eigenvectors of $\mathcal{H}_{HF}$.  The Hartree-field $E_{H}$ captures the 
electrostatic contribution to the bilayer capacitance, $E_{H} = (e^{2}/\varepsilon l_{B}) (d/2 l_B)$,
and the exchange fields capture fermion quantum-statistics: 
\begin{equation}
X^{\xi}_{n_{2}n'nn_{1}} = \int \, \frac{d^{2}\bf{p}}{(2\pi)^2} \, v_{\xi}({\bf p})  
F_{n_{2}n'} ({\bf p}) F_{nn_{1}} (-{\bf p}).
\label{exchange} 
\end{equation}
In Eq.(~\ref{exchange}) $v_{\pm}$ are the symmetric and antisymmetric combinations of same (s) 
and different (d) layer electron-electron interactions ($v_{s} = 2\pi 
e^{2}/\varepsilon q$ \, $v_{d}= v_{s}e^{-qd}$),
and the form factors ($F_{00} ({\bf q}) = e^{-(ql_{B})^{2}/4}$, $F_{10} ({\bf q}) =  
(i q_{x}+q_{y}) l_{B)} e^{-(ql_{B})^{2}/4}/\sqrt{2} = [{F}_{01}(-\bf{q})]^*$ and $ F_{11}({\bf q}) = 
(1-(ql_{B})^{2}/2)e^{(-ql_{B})^{2}/4}$) reflect the character of the two different quantum cyclotron orbits.


\begin{figure}[t]
\begin{center}
\includegraphics[clip,width=3.375in,height=2.25in]{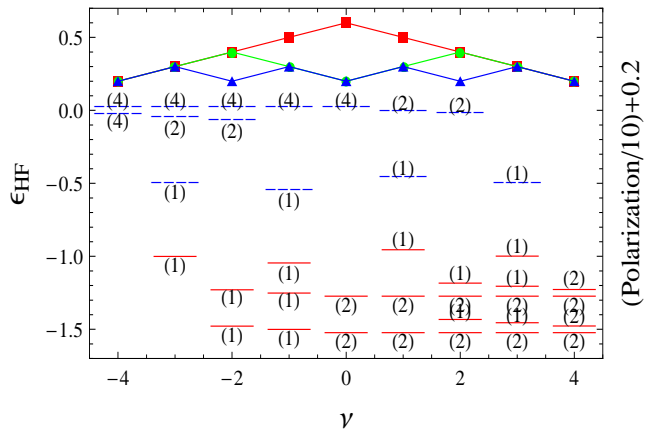}
\caption{(Color online) Filling factor dependence of the integer filling factor HF theory
occupied state ( spectrum 
of the bilayer graphene octet at $\Delta_{V}=0$).  Energies of occupied (red - solid lines) and 
unoccupied (blue - dashed lines) are in units of 
$(\pi/2)^{1/2} e^{2}/\varepsilon l_{B}$.  The Zeeman field $\Delta_{Z}$ value 
in these units is 0.023 at a magnetic field of $20 T$. 
Octet space fractional pseudospin polarizations offset for clarity: spin(red boxes), valley(green circles) and 
LL pseudospin(blue triangles). }
\label{energygaps}
\end{center}
\end{figure}

The solution of the Hartree-Fock equations for balanced bilayers ($\Delta_{V}=0$) is summarized in 
Fig.[~\ref{energygaps}] using a Zeeman field strength corresponding to $B = 20 $T.
The large gaps ($\sim (\pi/8)^{1/2}$ in $e^{2}/\varepsilon l_{B}$ units) 
separating occupied and empty states at the odd integer filling factors of primary interest  
justify our weak-coupling theory. 
The octet filling, proceeding in integer increments starting from filling factor $\nu=-4$, follows a 
Hunds rule behavior:  first maximize spin-polarization, then maximize layer-polarization to the 
greatest extent possible, then maximize Landau-level polarization to the extent allowed by the 
first two rules.  
For balanced bilayers  
the layer symmetric states (S) are filled before the layer antisymmetric states (AS). 
The first four states to be filled are (S,$n=0,\uparrow$),(S,$n=1,\uparrow$),(AS,$n=0,\uparrow$) and 
(AS,$n=1,\uparrow$) in this order.
This sequence is then repeated for the next four states with down ($\downarrow$) spin.
The Hunds rules imply that the Landau-level pseudospin is polarized at all odd integer 
filling factors between $\nu=-4$ and $\nu=4$.  The physics of this new type of pseudospin 
polarization is the main focus of this paper.  An important distinction between layer 
and Landau-level polarization is that the former is associated with spontaneous
inter-layer phase coherence whenever a Landau level occupies both layers simultaneously,
whereas the latter polarization is driven by the Landau-level dependence of the
microscopic Hamiltonian.  


Octet quantum Hall ferromagnets have an interesting and intricate 
dependence on the external potential $\Delta_{V}$.  Because the two-layers are close together, a small
value of $\Delta_{V}$ is sufficient to change the character of the layer polarization from the 
$XY$ spontaneous-coherence form, to an Ising polarization form in which one layer is occupied before the other.
We find that for $\Delta_{V} $ larger than a critical value $\Delta_{V}^*$, the layer  
filling proceeds by filling the top layer first. (For $\nu=-3$, $\Delta_{V}^* = 0.082(0.31)$ meV at $B= 20(50)$ 
Tesla.)  
As we explain later, this filling sequence has qualitative 
consequences for the odd-integer filling factor LL pseudospin polarized states.
\newline

\noindent 
{\em Landau-Level Pseudospin Dipoles}--- We now focus on the LL pseudospin fluctuations of a 
state with odd-integer filling factor, freezing spin and layer degrees of freedom.
The collective excitation 
spectrum of graphene bilayer octets as a function of $\nu$ and $\Delta_{V}$ will
be described in full detail elsewhere~\cite{longpaper}.
Fluctuating LL spinors are linear combinations of $n=0$ orbitals (even with respect to
their cyclotron orbit center) and $n=1$ orbitals (odd with respect to orbit center),
and therefore carry an electric dipole proportional to the in-plane component of their pseudospin.
Because dipole-dipole interactions are long-range, they play a dominant role in the 
QHF long-wavelength effective action\cite{moonprb}.  We find that   
\begin{equation}
\label{effaction}
S[\vec{m}] = \int dt \big[ \int d^{2}q \; \vec{\mathcal{A}} \cdot \partial_{t} \vec{m} - 
E[\vec{m}] \big] , 
\end{equation}
where the first term is the Berry-phase contribution\cite{moonprb,auerbach} and for 
small fluctuations away from $m_{z}=1$ (full $n=0$ polarization) 
\begin{equation}
\label{effene}
E[\vec{m}] = \frac{e^{2}}{\varepsilon l_{B}} \int d^{2}q  \big[ \frac{1}{2 |q|} (\vec{q} \cdot \vec{m} 
)^{2}+ \frac{\tilde{\Delta}_{LL}}{2} (m_{x}^{2} + m_{y}^{2}) \big] .
\end{equation} 
where $ \tilde{\Delta}_{LL}= \Delta_{LL}/(e^{2}/\epsilon l_{B})$. The mass terms in Eq.(~\ref{effene}) are due to the 
single-particle splitting between $n=0$ and $n=1$ 
levels and the interaction term is due to electric-dipole interactions.  The absence of 
interaction contributions to the mass terms is a surprise, since the interaction is Landau-level
pseudospin dependent.  We address this point below.   
Because of the in-plane electric dipoles associated with 
LL pseudospinors, the long-wavelength pseudo-spinwave collective mode dispersion is not analytic:
$\hbar \omega \to (\Delta_{LL}^{2} + \Delta_{LL} e^2q /\epsilon)^{1/2} $, 
and for $\Delta_{LL} \to 0$ is proportional to $q^{3/2}$ when exchange interactions are included in the energy functional.
The in-plane dipoles are also responsible for the intra-Landau-level cyclotron resonance discussed below.

To explain the absence of interaction contributions to the mass terms and address shorter-wavelength 
fluctuations it is necessary to derive the action microscopically.  It is convenient to 
temporarily restrict fluctuations to one space direction 
by considering Landau-gauge states in which the LL pseudospins
at different guiding centers $X$ fluctuate independently: 
\begin{equation}
|\psi [z] \rangle = \prod_{{\tiny X}} (z_{0 {\small X}} c^{\dagger}_{0 {\small X}} + 
z_{1{\small X}} c^{\dagger}_{1{\small X}}) |0 \rangle,
\end{equation}
where the spinor components $z_{n{\small X}}$ satisfy the normalization constraint
$|z_{0{\small X}}|^{2} + |z_{1{\small X}}|^{2} = 1$.
The corresponding imaginary-time action is 
\begin{widetext}
\begin{equation}
\label{action}
\mathcal{S}[\bar{z},z] = \mathcal{S}_{B}+\mathcal{E} = \int_{0}^{\beta} d \tau \sum_{{\small 
X} n} \bar{z}_{n {\small X}} \partial_{\tau} 
z_{n {\small X}} + \sum_{{\small X X'}} ( \frac{1}{2} \sum_{{n_{i}}} \big[ H (X-X') - F (X-X') 
\big] 
\bar{z}_{n_{1} {\small X}} z_{n_{3} {\small X}}
\bar{z}_{n_{2} {\small X'}} z_{n_{4}{\small X'}}  + \xi \Delta_{LL} \bar{z}_{1{\small 
X}} z_{1{\small X'}} ),
\end{equation}
\end{widetext}
where $\mathcal{S}_{B}$ is the Berry's phase term and 
$\mathcal{E}= \langle \psi [z]| ({\cal H} + {\cal H}_{int})|\psi [z] \rangle$
is the energy functional.
In Eq.~(\ref{action}) the direct($H$) and exchange($F$) 
energy contributions depend on the LL pseudospin labels,  
\begin{eqnarray}
H^{\,n_1,n_2}_{n_3,n_4}(X)  &=& \frac{1}{L_{y}} \int \frac{dq}{2 \pi} v_{q} F_{n_{1}n_{4}}(q)F_{n_{2}n_{3}}(-q) 
e^{-iq_{x}X}, \nonumber \\
F^{\,n_1,n_2}_{n_3,n_4}(X) &=& \frac{1}{L^{2}} \sum_{{\bf q}} v_{{\bf q}} \delta_{q_{y},X} F_{n_{1}n_{3}}({\bf 
q})F_{n_{2}n_{4}}(-{\bf q}).
\end{eqnarray} 
This action can be identified as the Schwinger boson\cite{auerbach} coherent state path integral representation of a 
model with pseudospins at each guiding center.
We can introduce a bosonic creation operator $a^{\dagger}_{n X} $ 
corresponding to $\bar{z}_{n X}$ and let $\mathcal{E}[\bar{z},z] \to 
\mathcal{H}[a^{\dagger},a] $.  

To analyze fluctuations around the HF mean field state, we use the linear spin 
wave approximation
\begin{equation} 
a_{0 X} \to 1 - \frac{1}{2} a^{\dagger}_{X} a_{X} \qquad a_{1,X} \to a_{X}.
\end{equation}
Taking the continuum limit $1/L_{y} \sum_{X} = \int dX/(2 \pi l_{B}) $, the action describing harmonic 
fluctuations can be written in Fourier space as  $S= S_{0} + \delta S$ where
\begin{equation}
\label{actionfluc}
\delta S =  \frac{e^{2}}{\varepsilon l_{B}} \int_{0}^{\beta} d \tau \sum_{{\bf q}} \bigg[ 
(\frac{1}{2} 
\sqrt{\frac{\pi}{2}} + \xi_{q} ) a^{\dagger}_{q} a_{q} + \frac{\lambda_{q}}{2} (a_{q}a_{-q} + 
a^{\dagger}_{q} a^{\dagger}_{-q}) \bigg],
\end{equation}
with
\begin{eqnarray} 
\nonumber
\xi_{q} &=&  \frac{|ql_{B}|}{2} e^{\frac{-(ql_{B})^{2}}{2}} -  \int dp  
\big( 1-\frac{p^{2}}{2} \big) J_{0} (ql_{B} p)e^{\frac{-p^{2}}{2}} + \xi \tilde{\Delta}_{LL}, \\
\lambda_{q} &=& \frac{|ql_{B}|}{2} e^{\frac{-(ql_{B})^{2}}{2}} -  \int dp
\frac{p^{2}}{2} J_{2} (ql_{B} p)e^{\frac{-p^{2}}{2}},
\end{eqnarray}
In Eq.(~\ref{actionfluc}) we have restored \cite{burkovahm} two-dimensional wavevectors to recognize the system's 
spatial anisotropy. 
The first and second terms in the above expressions capture the direct($H$) and  exchange($F$) 
contributions respectively and $J_{0}$ and $J_{2}$ are the zeroth and second order Bessel functions.  
The quadratic action in Eq.~(\ref{actionfluc}) has the familiar Bogoliubov form and the energy 
dispersion of the collective mode is given by:
\begin{equation} 
\omega(q) = \frac{e^{2}}{\varepsilon l_{B}} \big( (\frac{1}{2} \sqrt{\frac{\pi}{2}} + \xi_{q})^{2} - 
|\lambda_{q}|^{2} \big)^{1/2}. 
\end{equation} 
\begin{figure}[t] 
\begin{center} 
\includegraphics[clip,width=3.375in,height=2.25in]{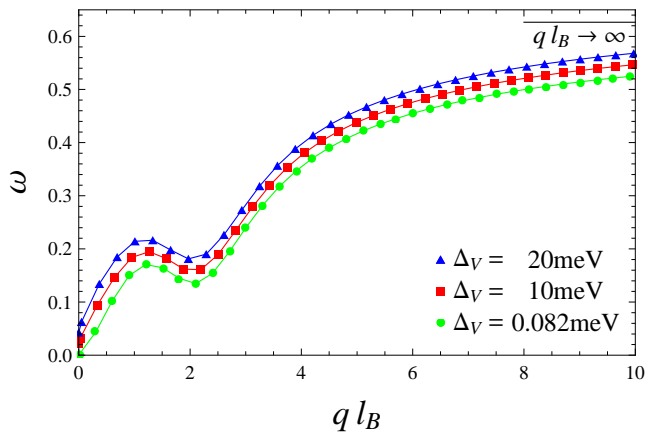} 
\caption{Collective mode $\omega_{q}$ of the Landau-level pseudospin polarized
state in units of interaction strength 
$e^{2}/\epsilon l_{B}=11.2\sqrt{B[Tesla]}$ meV
as a function of $q l_{B}$ at different values of the 
external potential difference $\Delta_{V}$ at a magnetic field of $20$ T. 
The black(solid) line indicates the $q l_{B} \to \infty $ asymptote for $\Delta_{B} = 0$.} 
\label{dispersion} 
\end{center} 
\end{figure}


As shown in Fig.[~\ref{dispersion}], this collective mode 
has a roton minimum at $ql_{B} \thickapprox 2.3$ and 
approaches the Hartree-Fock theory band splitting for $q \to \infty$
as expected.\cite{kallinhalperin1984} 
The surprising absence of interaction contributions to the gap at $q=0$ can be 
understood by examining the dependence of the 
uniform state interaction energy on global rotations in LL pseudospin space: 
\begin{equation}
\label{energy} 
\frac{2 \mathcal{E}[z]}{N_{\phi}} = - \frac{e^{2}}{\varepsilon l_{B}} 
\sqrt{\frac{\pi}{2}} \bigg[ |z_{0}|^{4}  + 
\frac{3}{4}|z_{1}|^{4} 
+ 2 |z_{0}|^{2} |z_{1}|^{2} \bigg], 
\end{equation}
The factor in square brackets above is $1 - |z_1|^{4}/4$, independent of $z_1$ to
quadratic order.  Notice that because $\Delta_{LL} <0$ for $\nu=-1,3$ the absence of 
interaction contributions to the gap implies that the fully spin-polarized state 
is unstable.  The ground state at these filling factors is instead\cite{longpaper} an $XY$ state with 
spontaneous phase order.  
\newline
\noindent 
{\em Intra-Landau-Level Cyclotron Resonance}---Finally we show that the octet QHF will exhibit unusual 
intra-LL cyclotron modes at odd filling factors, focusing on the fully 
polarized $\nu=-3,1$ cases.  The dynamical 
conductivity $\sigma_{\pm} = \sigma_{xx} \pm  i \sigma_{xy}$ can be evaluated using 
linear response theory. The projection of the current operator,  $j_{i} = d \mathcal{H}/d \pi_{i}$, 
onto the octet space can be expressed in terms of LL pseudospins:
\begin{equation}
j_{i} = \frac{\xi \Delta_{B}}{m \gamma_{1}} \big( \frac{\hbar}{\sqrt{2} l_{B}} m_{i} + 
\frac{e}{c}\mathcal{A}^{ext}_{i}(t) \big), 
\end{equation}
where the {\em ac} electric field $ {E}_{i} = (1/c) d \mathcal{A}^{ext}_{i}/ dt$.  The 
{\em ac} conductivity $(\xi = 1)$ is most simply evaluated by solving the LL pseudospin equation of 
motion with the $j \cdot \mathcal{A}^{ext}$ coupling included in the energy functional.  We find that  
\begin{equation}
\sigma_{\pm}(\omega) = \frac{N_{\phi}e \Delta_{B}}{m \gamma_{1}} \, \frac{1}{i(\omega \pm \,
\omega_{LL})}
\end{equation}
In the absence of interactions the conductivity has intra-octet peaks at the LL band-splitting frequency 
$\omega_{LL}$, in addition to inter-Landau-level peaks which do not appear in the projected
theory.  Trigonal warping is expected~\cite{graphenebilayertheory} to have little influence on 
electronic properties over the broad field range over which $\hbar l^{-1}_{B} > v_{3}m $. 
By performing explicit numerical calculations on the four-band model of bilayer graphene 
for typical bilayer quantum Hall parameters, a magnetic field strength of $10 {\rm Tesla}$ and 
$\Delta_{V} \approx 10 {\rm meV}$, we have verified that both the position 
and the oscillator strength of the intra-Landau level resonance are shifted by less than $2\%$. 

These low-frequency absorption peaks should be visible in microwave absorption experiments.
The appearance of tunable low-frequency peaks in $\sigma(\omega)$ is a surprise that might 
be quite interesting from the point of view of the quantum Hall localization physics, even in 
systems for which disorder dominates interactions.  In normal quantum Hall systems, peaks in 
$\sigma_{\pm}$ appear near the characteristic inter-Landau-level energy $\omega_c$ and the strong localization physics which leads to flat 
broad quantum Hall pleateaus occurs only in systems with $\omega_c \tau > 1$.  We conjecture that 
one requirement for odd-integer filling factor plateaus within the graphene bilayer octet 
is that $\omega_{LL} \tau > 1$.  Since $\omega_{LL}$ is proportional to $\Delta_{V}$, the strength of the 
quantum Hall effect can be tuned by a gate voltage which doesn't influence either the system's disorder 
or its total carrier density. \newline

This work was supported in part by the Welch Foundation, by the National Science Foundation under grant 
DMR-0606489 and by a research grant from the Natural Sciences and Engineering Research Council of
Canada (NSERC). Y.B. would like to acknowledge helpful discussions with Hongki Min.


\begin{thebibliography}{99}
\bibitem{trigonalfootnote} As shown in \onlinecite{graphenebilayerexpt}, the neglect of the trigonal
warping effect is justified in a sufficiently strong magnetic field.
\bibitem{valleylayercaveat} Since the K valley states in the octet are 
are localized in the the top (t) layer, and the K' valley states are 
localized in the bottom (b) layer, the valley (K,K') degree of freedom is 
equivalent to the layer (t,b) degree of freedom.
\end{thebibliography}
\end{document}